\documentclass[aps,twocolumn,showpacs,pra,twoside,amssymb,amsmath]{revtex4}
\usepackage{amssymb}
\usepackage{graphicx}
\usepackage{amsmath}
\usepackage{colordvi}
\usepackage{bbm}

\newcommand{\eins}{\ensuremath{\mathbbm 1}}

\begin{document}

\title{Observable estimation of entanglement for arbitrary finite-dimensional mixed states}
\author{Cheng-Jie Zhang}
\email{zhangcj@mail.ustc.edu.cn}
\author{Yan-Xiao Gong}
\author{Yong-Sheng Zhang}
\email{yshzhang@ustc.edu.cn}
\author{Guang-Can Guo}
\affiliation{Key Laboratory of Quantum Information, University of
Science and Technology of China, Hefei, Anhui 230026, People's
Republic of China}

\begin{abstract}
We present observable upper bounds of squared concurrence, which are
the dual inequalities of the observable lower bounds introduced in
[F. Mintert and A. Buchleitner, Phys. Rev. Lett. \textbf{98}, 140505
(2007)] and [L. Aolita, A. Buchleitner and F. Mintert, Phys. Rev. A
\textbf{78}, 022308 (2008)]. These bounds can be used to estimate
entanglement for arbitrary experimental unknown finite-dimensional
states by few experimental measurements on a twofold copy
$\rho\otimes\rho$ of the mixed states. Furthermore, the degree of
mixing for a mixed state and some properties of the linear entropy
also have certain relations with its upper and lower bounds of
squared concurrence.
\end{abstract}


\maketitle

\section{Introduction}
%\textit{Introduction.--}
Entanglement is not only one of the most fascinating features of
quantum theory that has puzzled generations of physicists, but also
an essential resource in quantum information
\cite{nielsen,Bell,QPT,werner}. Thus, the detection
\cite{detection1,detection2,detection3,detection4,detection5} and
quantification
\cite{review1,concurrence,concurrence1,hierarchy,Fan,Fei,Peres,tangle,Mintert}
of entanglement became fundamental problems in quantum information
science. A number of measures have been proposed to quantify
entanglement, such as concurrence \cite{concurrence,concurrence1},
negativity \cite{Peres} and tangle \cite{tangle}.

Recently, much interest has been focused on the experimental
quantification of entanglement
\cite{Mintert1,Mintert2,Mintert3,Mintert4,Mintert5,Eisert,otfried,Yu,Ren,nature,nature2,sun}.
On the one hand, original methods of experimentally detecting
entanglement are entanglement witnesses (EWs) \cite{witness1} which,
however, require some a priori knowledge on the state to be
detected. On the other hand, quantum state tomography needs rapidly
growing experimental resources as the dimensionality of the system
increases. To overcome shortcoming of EWs and the tomography,
Mintert \textit{et al.} proposed a method to directly measure
entanglement on a twofold copy $|\psi\rangle\otimes|\psi\rangle$ of
pure states \cite{Mintert1}. With this method, Refs.
\cite{nature,nature2} and \cite{sun} reported experimental
determination of concurrence for two-qubit and
$4\times4$-dimensional pure states, respectively. Moreover, Mintert
\textit{et al.} also presented observable lower bounds of squared
concurrence for arbitrary bipartite mixed states \cite{Mintert2} and
multipartite mixed states \cite{Mintert3}. For experimental
\textit{unknown} states, observable upper bounds of concurrence can
also provide an estimation of entanglement. Obviously, measuring
upper and lower bounds in experiments can present an exact region
which must contain the squared concurrence of experimental quantum
states.

The convex roof construction for mixed state concurrence indicates
that any direct decomposition of the state $\rho$ into pure states
will yield an upper bound of the entanglement of $\rho$. However,
for arbitrary experimental \textit{ unknown} mixed states, the
observable upper bound is non-trivial since it also provides an
estimation of entanglement as well as the lower bound in
experiments.

In this paper, we present observable upper bounds of squared
concurrence which, together with the observable lower bounds
introduced by Mintert \textit{et al.} \cite{Mintert2,Mintert3}, can
estimate entanglement for arbitrary experimental unknown states.
These bounds can be easily obtained by few experimental measurements
on a twofold copy $\rho\otimes\rho$ of the mixed states. Actually,
the upper bounds are the dual one of the lower bounds in Refs.
\cite{Mintert2,Mintert3}. Furthermore, the degree of mixing for a
mixed state and some properties of the linear entropy also have
certain relations with its upper and lower bounds of squared
concurrence.

The paper is organized as follows. In Sec. II we propose an
observable upper bound of squared concurrence for bipartite states
and multipartite states. The relations with properties of the linear
entropy is shown in Sec. III. In Sec. IV we discuss a tighter upper
bound of squared concurrence for two-qubit states, and give a brief
conclusion of our results.

\section{Observable upper bound for arbitrary mixed states}
\textit{Bipartite mixed states.} The \textit{I} concurrence of a
bipartite pure state is defined as \cite{concurrence1,Mintert1}
\begin{equation}\label{pure}
    C(|\psi\rangle)\equiv\sqrt{2(1-\mathrm{Tr}\rho_{A}^{2})}=
    \sqrt{\langle\psi|\otimes\langle\psi|A|\psi\rangle\otimes|\psi\rangle},
\end{equation}
where the reduced density matrix $\rho_{A}$ is obtained by tracing
over the subsystem B and $A=4P_{-}^{(1)}\otimes P_{-}^{(2)}$.
$P_{-}^{(i)}$ ($P_{+}^{(i)}$) is the projector on the antisymmetric
subspace $\mathcal{H}_{i}\wedge\mathcal{H}_{i}$ (symmetric subspace
$\mathcal{H}_{i}\odot\mathcal{H}_{i}$) of the two copies of the
$i$th subsystem $\mathcal{H}_{i}\otimes\mathcal{H}_{i}$, which has
been defined as follows \cite{Mintert1}
\begin{eqnarray}
P_{\mp}^{(i)}=\frac{1}{4}\sum_{jk}(|\alpha_{j}\alpha_{k}\rangle\mp|\alpha_{k}\alpha_{j}\rangle)(\langle\alpha_{j}\alpha_{k}|\mp\langle\alpha_{k}\alpha_{j}|),
\end{eqnarray}
where $\{|\alpha_{j}\rangle\}$ is an arbitrary complete set of
orthogonal bases of $\mathcal{H}_{i}$. The definition of \textit{I}
concurrence can be extended to mixed states $\rho$ by the convex
roof, $C(\rho)=\inf_{\{p_{i},|\psi_{i}\rangle\}}
\sum_{i}p_{i}C(|\psi_{i}\rangle),
    \   \rho=\sum_{i}p_{i}|\psi_{i}\rangle\langle\psi_{i}|$,
for all possible decomposition into pure states, where $p_{i}\geq0$
and $\sum_{i}p_{i}=1$. Ref. \cite{Mintert2} introduced lower bounds
of squared concurrence for arbitrary finite-dimensional bipartite
states,
\begin{equation}\label{}
   [C(\rho)]^{2}\geq\mathrm{Tr}(\rho\otimes\rho V_{i}),
\end{equation}
with $V_{1}=4(P_{-}^{(1)}-P_{+}^{(1)})\otimes P_{-}^{(2)}$ and
$V_{2}=4P_{-}^{(1)}\otimes(P_{-}^{(2)}-P_{+}^{(2)})$. We conjecture
its dual inequality as follows
\begin{equation}\label{bipartite}
[C(\rho)]^{2}\leq\mathrm{Tr}(\rho\otimes\rho K_{i}),
\end{equation}
with $K_{1}=4P_{-}^{(1)}\otimes \eins^{(2)}$ and
$K_{2}=4(\eins^{(1)}\otimes P_{-}^{(2)})$. The proof of this
inequality is shown in the following.
\begin{eqnarray}
[C(\rho)]^{2}&=&[\inf
\sum_{i}p_{i}C(|\psi_{i}\rangle)]^{2}\nonumber\\
&\leq&\inf
\sum_{i}[\sqrt{p_{i}}C(|\psi_{i}\rangle)]^{2}\cdot\sum_{i}(\sqrt{p_{i}})^{2}\nonumber\\
&=&\inf\sum_{i}2p_{i}(1-\mathrm{Tr}(\rho_{i}^{A})^{2})\nonumber\\
&\leq&2(1-\mathrm{Tr}\rho_{A}^{2})\nonumber\\
&=&\mathrm{Tr}(\rho\otimes\rho K_{1}),\nonumber
\end{eqnarray}
where
$\rho_{i}^{A}=\mathrm{Tr}_{B}|\psi_{i}\rangle\langle\psi_{i}|$. The
first inequality holds by applying the Cauchy-Schwarz inequality
\cite{rank2}, the second one, which has also been proved in Ref.
\cite{detection4}, holds due to the convex property of
$\mathrm{Tr}\rho_{A}^{2}$, and the last equality can be proved
directly using the definition of $P_{\mp}^{(i)}$. Similarly, one can
also obtain the inequality
$[C(\rho)]^{2}\leq\mathrm{Tr}(\rho\otimes\rho K_{2})$.

Similar to the lower bounds, inequality (\ref{bipartite}) implies
some interesting consequences:

(1) The upper bounds can be expressed in terms of the purities of
$\rho_{A}$ and $\rho_{B}$, i.e.,
\begin{eqnarray}\label{K}
\begin{split}
\mathrm{Tr}(\rho\otimes\rho K_{1})=2(1-\mathrm{Tr}\rho_{A}^{2}),\\
\mathrm{Tr}(\rho\otimes\rho K_{2})=2(1-\mathrm{Tr}\rho_{B}^{2}),
\end{split}
\end{eqnarray}
which coincide with Eq. (\ref{pure}) for pure state concurrence.
Notice that Ref. \cite{Mintert2} has introduced similar equations
for lower bounds $V_{1}$ and $V_{2}$,
\begin{eqnarray}\label{V}
\begin{split}
\mathrm{Tr}(\rho\otimes\rho V_{1})=2(\mathrm{Tr}\rho^{2}-\mathrm{Tr}\rho_{A}^{2}),\\
\mathrm{Tr}(\rho\otimes\rho
V_{2})=2(\mathrm{Tr}\rho^{2}-\mathrm{Tr}\rho_{B}^{2}).
\end{split}
\end{eqnarray}

(2) The upper bounds can be directly measured, since it is given in
terms of expectation values of $P_{-}$. It is a little different
from the experimental measurement of pure state concurrence
$4P_{-}^{(1)}\otimes P_{-}^{(2)}$. Notice that
$4P_{-}^{(1)}\otimes\eins^{(2)}=4P_{-}^{(1)}\otimes
P_{-}^{(2)}+4P_{-}^{(1)}\otimes P_{+}^{(2)}$. For pure state
$|\psi\rangle$,
$\langle\psi|\otimes\langle\psi|4P_{-}^{(1)}\otimes\eins^{(2)}|\psi\rangle\otimes|\psi\rangle=
\langle\psi|\otimes\langle\psi|4P_{-}^{(1)}\otimes
P_{-}^{(2)}|\psi\rangle\otimes|\psi\rangle$. Actually, Refs.
\cite{nature,nature2} and \cite{sun} measured their concurrence via
$4P_{-}^{(1)}\otimes\eins^{(2)}$ instead of $4P_{-}^{(1)}\otimes
P_{-}^{(2)}$. In this sense, they obtained an upper bound rather
than concurrence itself.

(3) Interestingly, it is worth noting that
\begin{equation}\label{minus}
\mathrm{Tr}(\rho\otimes\rho K_{i})-\mathrm{Tr}(\rho\otimes\rho
V_{i})=2(1-\mathrm{Tr}\rho^{2}),
\end{equation}
i.e., the degree of mixing can be easily calculated out based on the
upper and lower bounds.

\begin{figure}
\begin{center}
\includegraphics[scale=0.75]{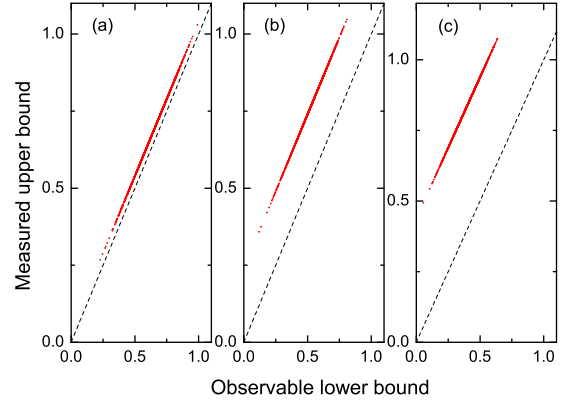}
\caption{(Color online). Measurable upper bound
$\mathrm{Tr}[\rho\otimes\rho(K_{1}+K_{2})/2]$ for squared
mixed-state concurrence $[C(\rho)]^{2}$, versus its observable lower
bound $\mathrm{Tr}[\rho\otimes\rho(V_{1}+V_{2})/2]$ for
$3\times3$-dimensional random states with different degrees of
mixing: \textit{a} shows weakly mixed states
($\mathrm{Tr}\rho^{2}=0.98$), \textit{b} displays intermediate
mixing ($\mathrm{Tr}\rho^{2}=0.88$), \textit{c} corresponds to
strongly mixed states ($\mathrm{Tr}\rho^{2}=0.78$). The dashed lines
denote the lower bound.}\label{1}
\end{center}
\end{figure}

Let us simulate the observable upper bound on mixed random states of
$3\times3$-dimensional systems. Mixed random states with different
degrees of mixing were obtained via the generalized depolarizing
channel \cite{channel}, as Ref. \cite{Mintert3} did. The observable
upper bound versus lower bound is shown in Fig. \ref{1}.
Interestingly, the upper bounds in Fig. \ref{1} are always in
parallel with the lower bounds, which actually coincides with Eq.
(\ref{minus}). For weakly mixed states, the bounds provide an
excellent estimation of concurrence; for strongly mixed states, they
also provide a region for concurrence.

\textit{Multipartite mixed states.} The generalized concurrence for
multipartite pure state is not unique. For instance, Ref.
\cite{Mintert1} introduced several inequivalent alternatives. In
this section, we choose the multipartite concurrence introduced in
\cite{multi,Mintert4}:
\begin{equation}\label{multipure}
C_{N}(\Psi)\equiv2^{1-N/2}\sqrt{(2^{N}-2)-\sum_{i}\mathrm{Tr}\rho_{i}^{2}},
\end{equation}
where $i$ labels all $(2^{N}-2)$ different reduced density matrices.
The definition can also be expressed as $C_{N}(\Psi)=
\sqrt{\langle\Psi|\otimes\langle\Psi|A|\Psi\rangle\otimes|\Psi\rangle}$
with $A=4(\mathbf{P}_{+}-P_{+}^{(1)}\otimes\cdots\otimes
P_{+}^{(N)})$. $\mathbf{P}_{+}$ ($\mathbf{P}_{-}$) is the projector
onto the globally symmetric (antisymmetric) space \cite{Mintert4}.
For mixed states, it is also given by the convex roof,
$C_{N}(\rho)=\inf_{\{p_{i},|\Psi_{i}\rangle\}}
\sum_{i}p_{i}C_{N}(\Psi_{i})$, for all possible decomposition into
pure states, where $p_{i}\geq0$ and $\sum_{i}p_{i}=1$. Ref.
\cite{Mintert3} introduced lower bounds of squared concurrence for
arbitrary multipartite states,
\begin{equation}\label{}
[C_{N}(\rho)]^{2}\geq\mathrm{Tr}(\rho\otimes\rho V),
\end{equation}
with $V=4(\mathbf{P}_{+}-P_{+}^{(1)}\otimes\cdots\otimes
P_{+}^{(N)}-(1-2^{1-N})\mathbf{P}_{-})$. We introduce an observable
$K$ such that
\begin{equation}\label{multipartite}
[C_{N}(\rho)]^{2}\leq\mathrm{Tr}(\rho\otimes\rho K),
\end{equation}
with $K=4(\mathbf{P}_{+}-P_{+}^{(1)}\otimes\cdots\otimes
P_{+}^{(N)}+(1-2^{1-N})\mathbf{P}_{-})$. The proof of this
inequality is shown in the following.
\begin{eqnarray}
[C_{N}(\rho)]^{2}&\leq&\inf\sum_{i}p_{i}[C_{N}(\Psi_{i})]^{2}\nonumber\\
&=&\inf\sum_{i}p_{i}2^{2-N}\sum_{k}c_{k}^{2}(\Psi_{i})\nonumber\\
&\leq&2^{2-N}\sum_{k}(2-\mathrm{Tr}\rho_{k}^{2}-\mathrm{Tr}\rho_{\overline{k}}^{2})\nonumber\\
&=&2^{2-N}\sum_{k}\mathrm{Tr}(\rho\otimes\rho\ 2(P_{-}^{(k)}\otimes
\eins+\eins\otimes P_{-}^{(\overline{k})}))\nonumber\\
&=&\mathrm{Tr}(\rho\otimes\rho K),\nonumber
\end{eqnarray}
where $\sum_{k}$ is taken over all the bipartite concurrence $c_{k}$
corresponding to each subdivision of the entire system into two
subsystems \cite{Mintert3}, $k$ denotes one subsystem and
$\overline{k}$ denotes the other one. We have used that
$\sum_{k}P_{-}^{(k)}\otimes
P_{-}^{(\overline{k})}=2^{N-2}(\mathbf{P}_{+}-P_{+}^{(1)}\otimes\cdots\otimes
P_{+}^{(N)})$ and $\sum_{k}(P_{-}^{(k)}\otimes
P_{+}^{(\overline{k})}+P_{+}^{(k)}\otimes
P_{-}^{(\overline{k})})=(2^{N-1}-1)\mathbf{P}_{-}$.

Inequality (\ref{multipartite}) also implies some interesting
consequences: (1) The upper bound can also be expressed in terms of
the purities of reduced density matrices, i.e.,
$\mathrm{Tr}(\rho\otimes\rho
K)=2^{2-N}[(2^{N}-2)-\sum_{i}\mathrm{Tr}\rho_{i}^{2}]$, which
coincides with Eq. (\ref{multipure}) for pure state concurrence. (2)
The upper bound can be directly measured, since it is given in terms
of expectation values of symmetric and antisymmetric projectors. It
is a little different from the lower bound
$\mathrm{Tr}(\rho\otimes\rho V)$. (3) Interestingly, it is worth
noting that
\begin{equation}\label{minus2}
\mathrm{Tr}(\rho\otimes\rho K)-\mathrm{Tr}(\rho\otimes\rho
V)=4(1-2^{1-N})(1-\mathrm{Tr}\rho^{2}),
\end{equation}
i.e., the degree of mixing can be easily calculated out based on the
upper and lower bounds.

\begin{figure}
\begin{center}
\includegraphics[scale=0.75]{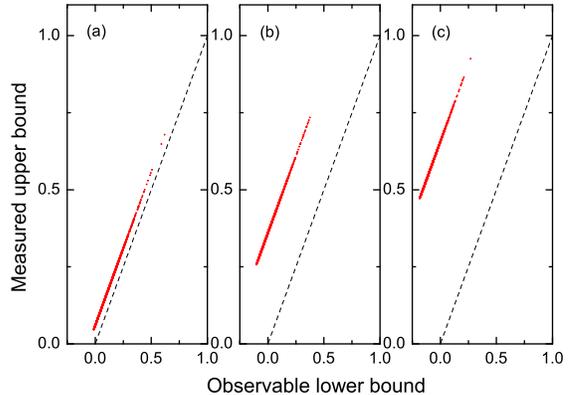}
\caption{(Color online). Measurable upper bound
$\mathrm{Tr}(\rho\otimes\rho K)$ versus its lower bound
$\mathrm{Tr}(\rho\otimes\rho V)$ for $2\times2\times2$-dimensional
mixed random states. Degrees of mixing are the same as Fig. \ref{1}.
}\label{2}
\end{center}
\end{figure}

We also simulate the observable upper bound on mixed random states
of $2\times2\times2$-dimensional systems with different degrees of
mixing obtained via the generalized depolarizing channel
\cite{channel}. The observable upper bound versus lower bound is
shown in Fig. \ref{2}. The upper bounds in Fig. \ref{2} are always
in parallel with the lower bounds as well, which actually coincides
with Eq. (\ref{minus2}). For weakly mixed states, the bounds provide
an excellent estimation of concurrence; for strongly mixed states,
they also provide a region for concurrence.

\section{Relations with properties of the linear entropy}
Interestingly, the upper and lower bounds of squared concurrence
have certain relations with some properties of the linear entropy,
such as the triangle inequality. The linear entropy is defined as
follows \cite{linearentropy}
\begin{equation}\label{}
    E(\rho)\equiv1-\mathrm{Tr}\ \rho^{2}.
\end{equation}
It can be regarded as a kind of linearized von Neumann entropy
$S(\rho)=-\mathrm{Tr}(\rho \log_{2} \rho)$, and has several same
properties as $S(\rho)$. In the following, we will give simple
proofs of the triangle inequality and subadditivity of the linear
entropy.

The triangle inequality can be proved directly using the upper and
lower bounds. Notice that the following inequalities hold for
arbitrary bipartite states:
\begin{eqnarray}
\begin{split}
\mathrm{Tr}(\rho\otimes\rho K_{1})\geq [C(\rho)]^{2}\geq
\mathrm{Tr}(\rho\otimes\rho V_{2}),\\
\mathrm{Tr}(\rho\otimes\rho K_{2})\geq [C(\rho)]^{2}\geq
\mathrm{Tr}(\rho\otimes\rho V_{1}).
\end{split}
\end{eqnarray}
Since Eqs. (\ref{K}) and (\ref{V}) hold, we can obtain some new
inequalities,
\begin{eqnarray}
\begin{split}\label{triangle1}
1-\mathrm{Tr}\rho_{A}^{2}\geq
\mathrm{Tr}\rho^{2}-\mathrm{Tr}\rho_{B}^{2},\\
1-\mathrm{Tr}\rho_{B}^{2}\geq
\mathrm{Tr}\rho^{2}-\mathrm{Tr}\rho_{A}^{2};
\end{split}
\end{eqnarray}
\begin{eqnarray}
\begin{split}\label{triangle2}
1-\mathrm{Tr}\rho^{2}\geq
(1-\mathrm{Tr}\rho_{B}^{2})-(1-\mathrm{Tr}\rho_{A}^{2}),\\
1-\mathrm{Tr}\rho^{2}\geq
(1-\mathrm{Tr}\rho_{A}^{2})-(1-\mathrm{Tr}\rho_{B}^{2}).
\end{split}
\end{eqnarray}
Obviously, inequalities (\ref{triangle2}) can be directly calculated
out from inequalities (\ref{triangle1}), and they are actually
triangle inequalities of the linear entropy,
\begin{eqnarray}
\begin{split}
E(\rho_{AB})\geq E(\rho_{B})-E(\rho_{A}),\\
E(\rho_{AB})\geq E(\rho_{A})-E(\rho_{B}).
\end{split}
\end{eqnarray}

Before embark on proving the subadditivity of the linear entropy,
let us review the universal state inverter $\widetilde{\rho}$
introduced in Ref. \cite{rank2},
\begin{eqnarray}
\widetilde{\rho}&\equiv&\mathrm{Tr}(\rho^{\dag})\eins\otimes\eins-\rho_{A}^{\dag}\otimes\eins-\eins\otimes\rho_{B}^{\dag}+\rho^{\dag}\nonumber\\
&=&\sum_{\alpha}\sigma_{y}\otimes\sigma_{y}(Q_{\alpha}\rho
Q_{\alpha})^{*}\sigma_{y}\otimes\sigma_{y},\nonumber
\end{eqnarray}
where $Q_{\alpha}=P_{A}^{(ii')}\otimes P_{B}^{(ii')}$,
$P_{A}^{(ii')}=|i\rangle_{A}\langle i|+|i'\rangle_{A}\langle i'|$
and $P_{B}^{(jj')}=|j\rangle_{B}\langle j|+|j'\rangle_{B}\langle
j'|$. The universal state inverter $\widetilde{\rho}$ is a
semi-positive definite operator, since each term in the sum
$\sigma_{y}\otimes\sigma_{y}(Q_{\alpha}\rho
Q_{\alpha})^{*}\sigma_{y}\otimes\sigma_{y}$ is semi-positive
definite. Therefore, $\sqrt{\rho}\widetilde{\rho}\sqrt{\rho}$ has
the semi-positive definite property as well, and we can obtain the
following inequality,
\begin{eqnarray}\label{17}
1+\mathrm{Tr}\rho^{2}-\mathrm{Tr}\rho_{A}^{2}-\mathrm{Tr}\rho_{B}^{2}
=\mathrm{Tr}\sqrt{\rho}\widetilde{\rho}\sqrt{\rho}\geq0,
\end{eqnarray}
where we have used
$\mathrm{Tr}\sqrt{\rho}\widetilde{\rho}\sqrt{\rho}=\mathrm{Tr}\rho\widetilde{\rho}=\mathrm{Tr}[\rho(\eins\otimes\eins-\rho_{A}\otimes\eins-\eins\otimes\rho_{B}+\rho)]=1+\mathrm{Tr}\rho^{2}-\mathrm{Tr}\rho_{A}^{2}-\mathrm{Tr}\rho_{B}^{2}$.
Thus,
$1+\mathrm{Tr}\rho^{2}-\mathrm{Tr}\rho_{A}^{2}-\mathrm{Tr}\rho_{B}^{2}\geq0$
holds, i.e. the subadditivity of the linear entropy
\begin{equation}\label{18}
    E(\rho_{A})+E(\rho_{B})\geq E(\rho_{AB})
\end{equation}
holds \cite{relation}.

In fact, it is not the first time to prove the subadditivity of the
linear entropy. For instance, Ref. \cite{Cai2} has proved the
subadditivity of the linear entropy. Furthermore, the triangle
inequality of the linear entropy can be proved from the
subadditivity \cite{Preskill}. Compared with this earlier proof,
roughly speaking, our proof is a little simpler. The main purpose of
this section is that these properties of the linear entropy are the
natural results from the positive semidefiniteness of the universal
state inverter and the upper and lower bounds, and they also
indicate the validity of these bounds.

\section{Discussions and conclusions}
Actually, for two-qubit states, $\mathrm{Tr}(\rho\otimes\rho\cdot 4
P_{-}^{(1)}\otimes P_{-}^{(2)})$ is a tighter upper bound of squared
concurrence than $\mathrm{Tr}(\rho\otimes\rho K)$. Because the
equation
\begin{equation}\label{pminus}
\mathrm{Tr}(\rho\otimes\rho\cdot 4P_{-}^{(1)}\otimes
P_{-}^{(2)})=\mathrm{Tr}\rho\widetilde{\rho}
\end{equation}
holds for arbitrary two-qubit states, where
$\widetilde{\rho}=\sigma_{y}\otimes\sigma_{y}\rho^{*}\sigma_{y}\otimes\sigma_{y}$.
Eq. (\ref{pminus}) has also been proved in \cite{Cai}. Furthermore,
notice that
$C=max\{\lambda_{1}-\lambda_{2}-\lambda_{3}-\lambda_{4},0\}$, where
$\lambda_{1},\lambda_{2},\lambda_{3},\lambda_{4}$ are squared roots
of eigenvalues of $\rho\widetilde{\rho}$ in the decreasing order.
Therefore, it is easily concluded that
$\mathrm{Tr}(\rho\otimes\rho\cdot4 P_{-}^{(1)}\otimes
P_{-}^{(2)})=\mathrm{Tr}\rho\widetilde{\rho}=\sum_{i}\lambda_{i}^{2}\geq(\lambda_{1}-\lambda_{2}-\lambda_{3}-\lambda_{4})^{2}\geq
C^{2}$. However, the new upper bound $\mathrm{Tr}(\rho\otimes\rho
\cdot4P_{-}^{(1)}\otimes P_{-}^{(2)})$ is hard to generalize to
arbitrary finite-dimensional bipartite states.

We give a brief discussion on the experimental measurement of our
upper bound. As only the projector $P_{-}$ on one of the subsystems,
rather than a complete set of observables, is required, our upper
bound could be easily measured. In particular, for two-dimensional
systems, $P_{-}$ is simply the projector onto the singlet state
\mbox{$|\Psi^{-}\rangle=(|01\rangle-|10\rangle)/\sqrt{2}$}. Let us
take the photonic system for example. The simplest way to project
two photons onto the singlet state is using a Hong-Ou-Mandel
interferometer \cite{PRL592044}. This method has been widely used
since the teleportation \cite{nature390575} experiment. Another
method, employed in \cite{nature,nature2}, is distinguishing the
Bell states with a controlled-NOT gate, which can transform the Bell
states to separable states \cite{PRL744083}.

In conclusion, we present observable upper bounds of squared
concurrence, which are the dual bound of the observable lower bounds
introduced by Mintert \textit{et al.}. These bounds can estimate
entanglement for arbitrary finite-dimensional experimental unknown
states by few experimental measurements on a twofold copy
$\rho\otimes\rho$ of the mixed states. Furthermore, the degree of
mixing for a mixed state and some properties of the linear entropy
also have certain relations with its upper and lower bounds of
squared concurrence. Last but not least, we discuss a tighter upper
bound for two-qubit states only, and it remains an open question to
generalize it to arbitrary finite-dimensional bipartite systems.

\section*{ACKNOWLEDGMENTS}
This work was funded by the National Fundamental Research Program
(Grant No. 2006CB921900), the National Natural Science Foundation of
China (Grants No. 10674127 and No. 60621064), the Innovation Funds
from the Chinese Academy of Sciences, Program for NCET.


\begin{thebibliography}{99}
\bibitem{relation} Actually, it is also held that
$\mathrm{Tr}(\rho\otimes\rho\cdot 4P_{-}^{(1)}\otimes
P_{-}^{(2)})=\mathrm{Tr}\sqrt{\rho}\widetilde{\rho}\sqrt{\rho}$,
since we have $\mathrm{Tr}(\rho\otimes\rho\cdot 4P_{-}^{(1)}\otimes
P_{-}^{(2)})=\mathrm{Tr}[\rho\otimes\rho(V_{1}+K_{2})/2]=\mathrm{Tr}[\rho\otimes\rho(V_{2}+K_{1})/2]=1+\mathrm{Tr}\rho^{2}-\mathrm{Tr}\rho_{A}^{2}-\mathrm{Tr}\rho_{B}^{2}$.
Therefore, the proof of the subadditivity above has a little
relations with the upper and lower bounds. Eq. (\ref{17}) indicates
that the average of the upper and lower bounds
$\mathrm{Tr}[\rho\otimes\rho(V_{1}+K_{2})/2]$ is nonnegative.
\end{thebibliography}
\end{document}